\documentclass[twocolumn,amsmath,amssymb]{revtex4-1}
\usepackage{graphicx}% Include figure files
\usepackage{dcolumn}% Align table columns on decimal point
\usepackage{bm}% bold math
\usepackage{color,soul}
\begin{document}

\title{Effective tuning of unusual Aharonov-Bohm oscillations in a single quantum ring}

\author{M.G. Barseghyan$^{1,2}$\thanks{mbarsegh@ysu.am},
A.Kh. Manaselyan$^1$, D. Laroze$^{3}$, A.A. Kirakosyan$^1$}

\affiliation{$^1$ Department of Solid State Physics, Yerevan State University, Yerevan, Armenia.}
\affiliation{$^2$ National University of Architecture and Construction of Armenia, Yerevan, Armenia.}
\affiliation{$^3$ Instituto de Alta Investigaci\'{o}n, CEDENNA, Universidad de Tarapac\'{a}, Casilla 7D, Arica, Chile.}
\affiliation{E-mail: manuk.barseghyan@nuaca.am (mbarsegh@ysu.am), amanasel@ysu.am, dlarozen@uta.cl, kirakosyan@ysu.am}

\begin{abstract}

The simultaneous effect of intense THz laser and in-plane electric fields on Aharonov-Bohm effect in single isotropic quantum rings is investigated using the non-perturbative Floquet theory in high-frequency limit. It is shown that in isotropic quantum rings the intense THz laser and in-plane electric field create unusual Aharonov-Bohm oscillations. For fixed values of intense THz laser field parameter, the amplitudes of Aharonov-Bohm oscillations can be effectively tuned by changing the electric field direction. Furthermore, for fixed values of electric field strength and laser field parameter the intraband optical properties can be effectively tuned by changing the electric field direction. Thus, it can be argued that a circular QR in the intense THz laser field is equivalent to the anisotropic quantum ring, and the electric field direction effectively tunes the unusual Aharonov-Bohm oscillations and intraband optical properties of the ring.

\end{abstract}

\maketitle

\section{Introduction}

Technologies involving nanometer-scale objects continue to improve the quality of our daily lives because the down-sizing of functional units can result in a significant decrease in device energy consumption and more efficient production processes. Therefore, the fabrication and investigation of nanometer-scale systems are undoubtedly one of the central issues in current science and technology \cite{Alivisatos}. A significant class of such nanometer-scale systems is quantum rings (QRs) \cite{Fomin1}, which attract special attention due to their fascinating physical properties.

On the other hand, one of the best-known effects that can be used to observe and control quantum interference is the Aharonov-Bohm (AB) effect \cite{Aharonov,Webb}. The AB oscillations for semiconductor QRs was experimentally observed and studied in Ref.~\cite{Schuster,Lorke,Fuhrer}. The introduction of new 2D materials, such as graphene \cite{Novesolelov}  and phosphorene \cite{Li} opened new ways to explore electronic and phase coherent transports in a two-dimensional system. Nanoscale QRs have been fabricated from graphene and phosforene, in which the phase coherence length at low temperatures is larger than or comparable to their circumferences. This implies that the AB effect can be observed in such QRs. Many experimental studies of the AB effect in graphene QRs have been made  \cite{Russo,Smirnov,Cabosart} but, to our knowledge, little on phosphorene QRs.

Theoretical investigation of the AB effect in different types of QRs has made great strides in recent years in unraveling new phenomena and their enormous potentials in device applications. The AB effect and persistent currents have been extensively studied in semiconductor QRs \cite{Wendler,Chakraborty,Splettstoesser,Fomin2}, and are expected to have potential applications in quantum electronics and quantum information. The influence of an in-plane electric field and eccentricity on the AB effect of a semiconductor QRs have been investigated in Ref.~\cite{Barticevic,Bruno}. It is shown that the electric field and the eccentricity may suppress the AB oscillations of the lower energy levels. The influence of lateral asymmetry on the electronic and optical properties in elliptical strained InAs QRs is analyzed in the presence of a perpendicular magnetic field \cite{Milosevic}. The calculations indicate that the AB oscillations in the energy of the ground state of elongated rings disappear for a large enough magnetic field. Recently, the effect of effective mass anisotropy, ellipticity of the confinement potential and electric field on AB oscillations in QRs have been investigated in Ref.~\cite{Sousa}. The usual AB oscillations are not observed for a circularly symmetric confinement potential. However, they can be reinstated if an elliptic QR is chosen. The influence of anisotropy, caused by the electron-electron interaction, on AB effect in new materials, such as ZnO QRs and dot-ring structures, was reported in Ref.~\cite{Chakraborty1,Chakraborty2}. The results indicate that the AB oscillations strongly depend on the electron number in that structures.

New studies on AB effect in graphene QRs have been made in Ref.~\cite{Abergel,Zarenia,Romanovsky}. Most recently, the AB effect in square phosphorene QRs, with armchair and zigzag edges, has been investigated in Ref.~\cite{Li1}. The main result obtained in this work is that the AB oscillation, observed in the energy spectrum, strongly depends on the ring width, electric field, and the side-gating potential. In our previous work, we have reported about theoretical investigations of THz intense laser field (ILF) effect on electronic and optical properties of isotropic and anisotropic GaAs QRs in the presence of magnetic field Ref.~\cite{Chakraborty3}. We have shown that in isotropic QRs the laser field creates the unusual AB oscillations. In the case of anisotropic QRs we have shown that with the ILF it is possible to completely control the anisotropy of the QR and thus the physical characteristics.

In aim of this article is to report theoretical investigations of the simultaneous influence of a THz intense laser field and lateral electric field on AB effect in isotropic QRs. The results have shown that the electric field destroys the AB oscillations in QRs even in the absence of ILF. It should be noted that this influence on energy levels is stronger for larger values of QR's width. Additionally, we have shown for the first time that the amplitude of unusual AB oscillations can be strongly controlled by the electric field direction. We have also shown, that by changing the in-plane electric field direction and laser field we can effectively control the unusual AB oscillations which are observed in intraband optical absorption. Therefore the simultaneous influence of in-plane electric field and ILF can effectively tune the unusual AB oscillations observed in energy and intraband optical spectra of isotropic QRs. TThe manuscript is organized as follow: In Sec. \ref{SII} the theoretical model is presented. In Sec. \ref{SIII} the numerical results are shown and analyzed. Finally, the conclusions are given in Sec. \ref{SIV}.

\section{Theoretical Framework}
\label{SII}

Our system consists of a two-dimensional isotropic QR structure containing single electron, under the action of a homogeneous in-plane electric field and magnetic field oriented along the growth direction. The system is radiated by ILF which is represented by a monochromatic plane wave of frequency $\nu$. The laser beam is non-resonant with the semiconductor structure and linearly polarized along a radial direction of the structure (chosen along the $x$-axis). The electron motion is described by the solution of the time-dependent Schr\"{o}dinger equation:
\begin{align}
\left[\frac1{2m}\left(\widehat{\textbf{p}}-\frac{e}{c}(\textbf{A}(t)+\textbf{A}_{m})
\right)^{2}+V(x,y)-e \textbf{F}\textbf{r}\right]\Phi(x,y,t) \nonumber\\
 =i\hbar\frac{\partial}{\partial t}\Phi(x,y,t)\,,
\end{align}
where $m$ is the effective mass, $e$ is the electron charge, $\widehat{\textbf{p}}$ is the lateral momentum operator of the electron, $\textbf{A}(t)=\textbf{e}^{}_xA^{}_0\cos(2\pi \nu t)$ is the laser field vector potential, where $\textbf{e}^{}_x$ denotes the unit vector on the $x$-axis, $\textbf{F}$ is the electric field strength. In Eq.~(1), $\textbf{A}^{}_m$ is the vector potential of the magnetic field which is chosen to be $\textbf{A}^{}_m=(0,Bx,0)$. In this case the scalar product $(\textbf{A}(t)\cdot\textbf{A}^{}_m)=0$. For the confinement potential $V(x,y)$ we choose the model of finite, square-well
type
\begin{equation}
V(x,y) = \left\{ \begin{array}{ll}
      0, & \mbox{if}\, \,R^{}_1\leq \sqrt{x^2+y^{2}} \leq R^{}_2,\\
      V^{}_0, & {\rm otherwise},\end{array}\right.
\end{equation}
where $R^{}_1$ and $R^{}_2$ are the inner and outer radii of the QR, respectively.

Using the dipole approximation and Kramers-Hennerberger unitary transformation \cite{Kramers,Henneberger} in the high-frequency regime the laser-dressed energies of the QR can be obtained from the following time-independent Schr\"{o}dinger equation \cite{Gavrila1,Pont,Valadares,Gavrila2,Ganichev,Baghramyan1,Baghramyan2}:
\begin{align}\label{static}
\left[\frac1{2m}\left(\widehat{\textbf{p}}-\frac{e}{c}\textbf{A}^{}_m\right)^2+V^{}_d(x,y)-e \textbf{F}\textbf{r}
\right]\Phi^{}_d(x,y) \nonumber\\
     =E^{}_d\Phi^{}_d(x,y)\,,
\end{align}
where $V^{}_d(x,y)$ is the time-averaged laser-dressed potential that can be expressed by
the following expression \cite{Radu1,Radu2,Chakraborty3}
\begin{align}\label{Vd}
V^{}_d(x,y)&=\frac{V^{}_0}{\pi}Re\biggl[\pi- \theta\left(\alpha_{0}-x-\Gamma^{}_1
\right)\arccos\left(\frac{\Gamma^{}_1+x}{\alpha^{}_0}\right) + \nonumber\\
           & + \theta\left(\alpha^{}_0-x -\Gamma^{}_2\right)\arccos\left(
\frac{\Gamma^{}_2+x }{\alpha^{}_0}\right)-\nonumber\\
           & - \theta\left(\alpha^{}_0+x -\Gamma^{}_1\right)\arccos\left(\frac{
\Gamma^{}_1-x }{\alpha^{}_0}\right)+\nonumber\\
           & + \theta\left(\alpha^{}_0+x -\Gamma^{}_2\right)\arccos\left(\frac{
\Gamma^{}_2-x }{\alpha^{}_0}\right)           \biggr] \,.
\end{align}
In Eq.(\ref{Vd}) $\Gamma^{}_i= Re\left(\sqrt{R_i^2-y^2}\right)$, $\theta(u)$ is the Heaviside unit-step function and $\alpha^{}_0=\left(e^{2}I/2{\pi}^{3}m^{2} \epsilon_h^{1/2}c \nu^{4}\right)^{1/2}$, which describes the strength of the laser field and comprises the intensity $I$, the frequency $\nu$ of the laser field, and the high-frequency dielectric constant $\epsilon^{}_h$. Note that $I$ and $\nu$ can be chosen for a broad range in units
of $kW/cm^2$  and terahertz, correspondingly \cite{Ganichev}.

 The eigenvalues of the laser dressed energy $E^{}_d$ and eigenfunctions
$\Phi^{}_d(x,y)$ may be obtained by solving Eq.~(\ref{static}) with the help of the exact diagonalization technique. The eigenfunctions are presented as a linear expansion of the eigenfunctions of the two-dimensional rectangular infinitely high potential well with radius $R>>R_2$ \cite{Chakraborty3,Gangopadhyay,Radu1,Radu2}. In our calculations we have used 361 basis states which are appropriate for determining the energy eigenvalues of ground and few excited states.

We have also considered here the intraband optical transitions in the conduction band. For $x$-polarization of the incident light the intensity of absorption
in the dipole approximation is proportional to the square of the dipole matrix element $M_{fi}\sim\langle
f|x|i\rangle$, when the transition goes from the initial state $|i\rangle$ to the final state
$|f\rangle$. In this work we always consider $|i\rangle$ to be the ground state.

\section{Results and discussion}
\label{SIII}

In this section we present the numerical simulations of our theoretical model. They are carried out for $GaAs$ QRs having parameters $V^{}_0=228$meV,
$m=0.067m^{}_0$ ($m^{}_0$ is the free electron mass) and $\epsilon^{}_h=10.9$ \cite{Adachi}.
The inner and outer radii of QR are $R^{}_1=5$ nm and $R^{}_2=20$ nm.

\vspace{0.3cm}
\begin{figure}
    \centering
 \resizebox{0.8\columnwidth}{!}{\includegraphics[width=0.35\textwidth,angle=0]{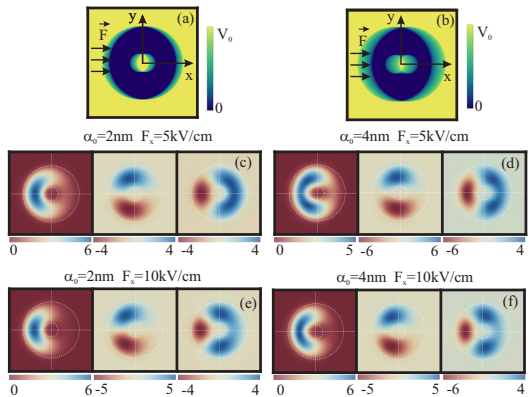}}
    \caption{\label{WFFx} (a) and (b): density plot of the dressed confinement potential for two values of the ILF parameter $\alpha^{}_0=2$nm and $\alpha^{}_0=4$nm. The arrows demonstrate direction of the electric field. (c)-(f): the density plots of the wave functions of first three states for different values of electric field $F$ and ILF parameter  $\alpha^{}_0$. The results are for $B=0$.}
\end{figure}
\vspace{0.3cm}

In the frames (a) and (b) of Fig.~\ref{WFFx}  is presented the schematic picture of the dressed confinement potential $V_d(x,y)$ for different values of the laser field intensity when the electric field is along $x$-axis. The ILF applied on a QR creates an anisotropy in the confinement potential as a result of which the effective length of the confinement along the x-direction decreases in the lower part of the QR potential well. The effects of ILF and electric field on the corresponding ground and first two excited states wave functions are presented in the frames (c) -(f) of  Fig.~\ref{WFFx}. As one can see, the electric field always moves the wave function to its opposite direction, but the strengthening of the ILF weakens the effect of the electric field due to the enhancement of the influence of size quantization for low laying states.

\vspace{0.3cm}
\begin{figure}
    \centering
 \resizebox{0.99\columnwidth}{!}{\includegraphics[width=0.35\textwidth,angle=0]{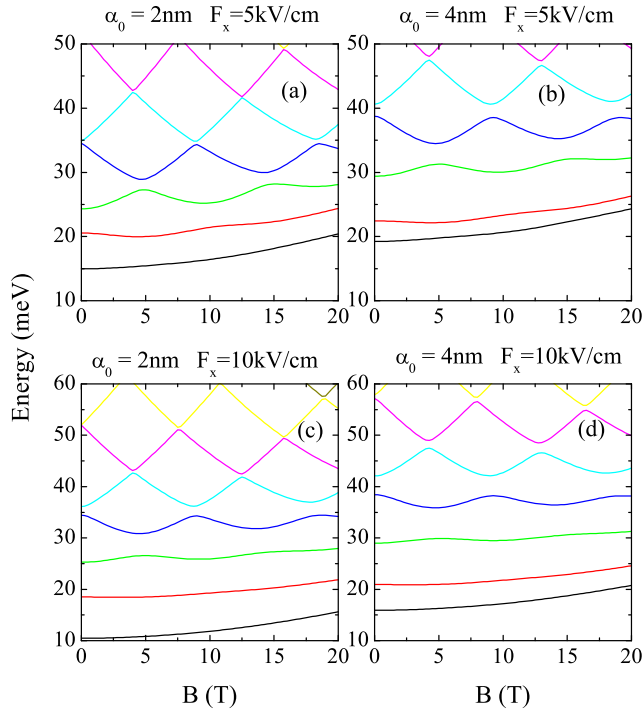}}
    \caption{\label{EFx} The dependences of low-lying energy levels of an electron on the magnetic field $B$ for different values of electric field $F_{x}$ and ILF parameter $\alpha^{}_0$.}
\end{figure}
\vspace{0.3cm}

Fig.~\ref{EFx} shows the magnetic field dependences of low laying energy levels of an electron for various values of the electric field, directed along $x$-axis, and ILF parameter $\alpha_0$. As it has been shown in our earlier work \cite{Chakraborty3}, the ILF creates an anisotropy in the confinement potential. Consequently, the degeneracy of the excited states at $B=0$ disappears. With an increase of $\alpha^{}_0$ due to the reducing symmetry from $C^{}_{\infty}$ to $C^{}_2$, one should expect an energy levels split into non-crossing pairs of states which in turn cross repeatedly as $B$ increases (unusual AB oscillations). On the other, hand the in-plane electric field creates polar symmetry and also destroys the AB oscillations in QR \cite{Barticevic}. Fig.~\ref{EFx} represents the combined effect of electric field and ILF on energy spectra and AB oscillations of QR. For the fixed value of $\alpha_0$ with the increase of $F_x$ the amplitude of unusual AB oscillations decreases. Similar effects have been previously observed for phosphorene QRs under applied electric field directed along $x$-axis \cite{Li1}.

\vspace{0.3cm}
\begin{figure}
    \centering
 \resizebox{0.8\columnwidth}{!}{\includegraphics[width=0.35\textwidth,angle=0]{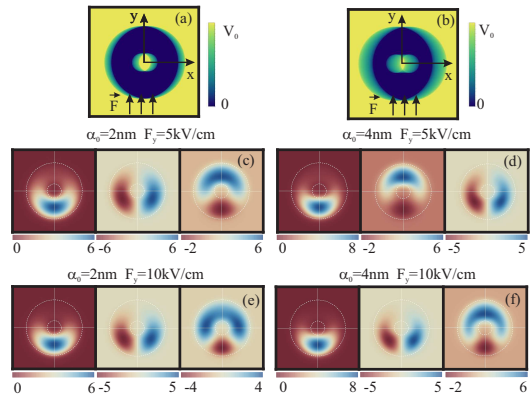}}
    \caption{\label{WFFy}Same as in FIG.1 but electric field is directed along the $y$-axis.}
\end{figure}
\vspace{0.3cm}

In the frames (a) and (b) of Fig.~\ref{WFFy} is presented the schematic picture of dressed confinement potential for different values of the laser field intensity when the electric field is along $y$-axis. The effects of ILF and electric field on the corresponding ground and first two excited states wave functions are presented in the frames (c) -(f) of  Fig.~\ref{WFFy}.  An interesting effect has been observed in Fig.~\ref{WFFy} (d) for the wave functions of the excited states. When $\alpha^{}_0=4$ nm and $F_y=5$ kV/cm the symmetry of the wave functions of excited states is changed due to the linear Stark effect in energy spectrum caused by the anisotropic modification of laser-dressed confinement potential \cite{Baghramyan1}. In this case, a crossing of energy levels is observed, as a result of which the order of energy levels and symmetry of wave functions of first and second excited states are changed.

\vspace{0.3cm}
\begin{figure}
    \centering
 \resizebox{0.98\columnwidth}{!}{\includegraphics[width=0.35\textwidth,angle=0]{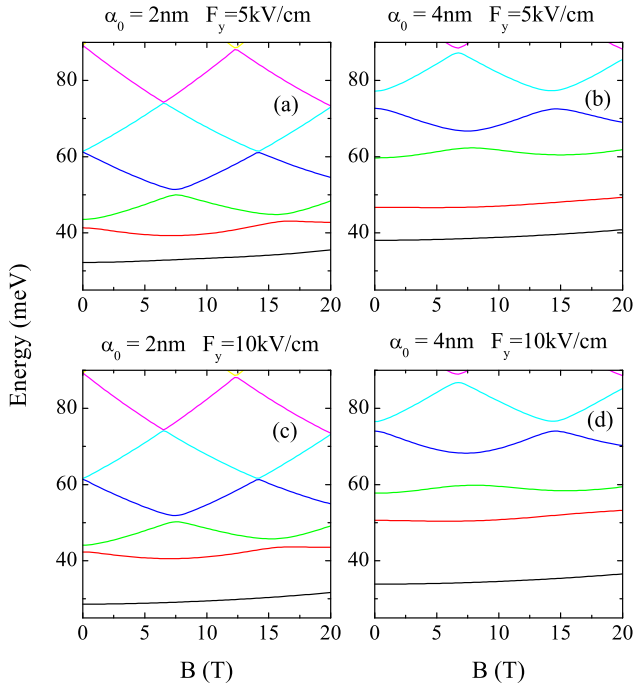}}
    \caption{\label{EFy}Same as in FIG.2 but electric field is directed along the $y$-axis.}
\end{figure}
\vspace{0.3cm}

Fig.~\ref{EFy} shows the magnetic field dependences of low laying energy levels for various values of the electric field, directed along $y$-axis, and ILF parameter $\alpha_0$. Here, the unusual AB oscillations under the simultaneous effect of electric field and ILF on energy levels are also observed. Comparing Fig.~\ref{EFy} (a)-(d) with Fig.~\ref{EFx} (a)-(d) we can see, that the amplitudes of unusual AB oscillations are much bigger when the electric field is directed along $y$-axis. It means that for the fixed values of $\alpha_0$ (for the fixed anisotropy of QR) the amplitudes of unusual AB oscillations can be increased by changing the direction of electric field from $x$-axis to $y$-axis.

\vspace{0.3cm}
\begin{figure}
    \centering
 \resizebox{0.9\columnwidth}{!}{\includegraphics[width=0.35\textwidth,angle=0]{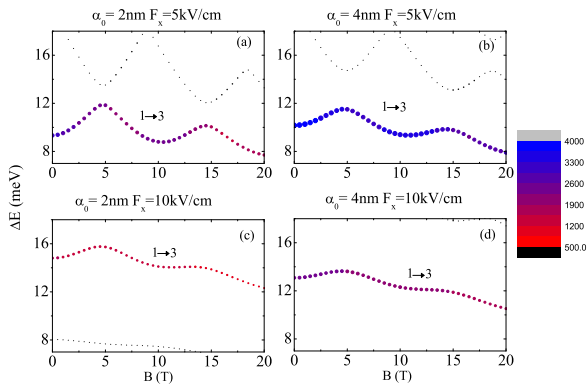}}
    \caption{\label{OptFx}The dependences of dipole allowed optical transition energies on magnetic field $B$
for different values of electric field $F_{x}$ and ILF parameter $\alpha^{}_0$. The size and the color of the circles are proportional to the intensity of the calculated optical transitions.}
\end{figure}
\vspace{0.3cm}

These unusual behaviors of electron wave functions and energy levels are expected to influence the intraband optical properties of the single QRs. Fig.~\ref{OptFx} shows the dipole-allowed intraband optical transition energies as a function of the magnetic field for different values of electric field strength and $\alpha_0$ when the electric field is directed along $x$-axis. The size and the color of the circles in this figure are proportional to the intensity $|M_{fi}|^{2}$ of the calculated optical transitions. As it can be seen from figures, for fixed values of electric field strength $F$ and ILF parameter $\alpha_0$ the unusual optical AB oscillations are observed. Note that for all cases the intensity of the $1\rightarrow3$ transitions (transitions from ground state to the second excited state)  more significant than others. This fact is caused by the symmetry of the corresponding wave functions and, as a result, by the value of corresponding dipole matrix element.

\vspace{0.3cm}
\begin{figure}
    \centering
 \resizebox{0.9\columnwidth}{!}{\includegraphics[width=0.35\textwidth,angle=0]{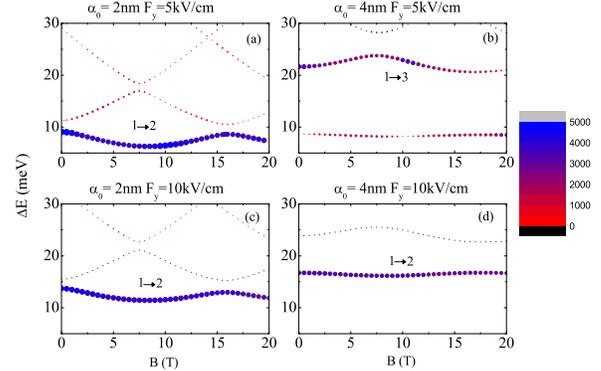}}
    \caption{\label{OptFy} Same as in FIG.5 but electric field is directed along the $y$-axis.}
\end{figure}
\vspace{0.3cm}

When the electric field is along $y$-axis one can also calculate the dipole-allowed intraband optical transition energies as it is displayed in Fig.~\ref{OptFy}. In this case, the unusual optical AB oscillations are also visible. Except of the Fig.~\ref{OptFy}(b) the intensity of the $1\rightarrow2$ transitions (transitions from ground state to the first excited state) much bigger than others. When $\alpha_0=4$ nm and  $F_y=5$ kV/cm the symmetry of the final state wave function changes, and $1\rightarrow3$ transitions become much stronger.

\section{Conclusions}
\label{SIV}

In conclusion, we have investigated here the simultaneous influence of an intense THz laser and in-plane electric field on the electronic and intraband optical properties of isotropic QRs in an applied magnetic field. We have shown that the simultaneous action of an intense THz laser and in-plane electric field creates the unusual AB oscillations. Additionally, we have found, that for fixed values of intense THz laser field parameter, the amplitudes of unusual Aharonov-Bohm oscillations can be effectively tuned by the changing of the electric field direction. Furthermore, we have shown that for fixed values of electric field strength and laser field parameter the intraband optical properties can be drastically influenced by the change of the electric field direction. Thus, it can be argued that for a fixed value of intense THz laser field parameter, which is equivalent to the anisotropic quantum ring, the changing of the electric field direction effectively tunes the unusual Aharonov-Bohm oscillations in energy spectra and intraband optical properties.  Lastly, it is worth noting that with the change of electric field direction the AB oscillations can become more regular. We believe that the results are useful and will open up new possibilities to improve the design and characterization of new devices based on single QR, such as THz detectors, efficient solar cells as well as photon emitters.

\section*{Acknowledgments}
DL acknowledges partial financial support from Centers of excellence with BASAL/CONICYT financing, Grant FB0807, CONICYT- ANILLO ACT 1410 and FONDECYT 1180905.

\end{document}